\documentstyle[12pt,aasms4]{article}
\newcommand{\al}{et\ al.\ }
\newcommand{\au}{\author}

\newcommand{\bib}{\bibitem}
\newcommand{\Eq}{equation\ }
\newcommand{\kb}{Kuiper\ belt\ }
\newcommand{\ara}{$a,\ R,\ \alpha$\ }
\begin{document}
\title{IR Kuiper Belt Constraints}

\au{Vigdor L. Teplitz,\altaffilmark{1}\altaffiltext{1}
{Physics Department, Southern Methodist University, Dallas, TX 75275}\
S. Alan Stern,\altaffilmark{2}\altaffiltext{2}
{Southwest Research Institute, Boulder, CO 80302}\
John D. Anderson,\altaffilmark{3}\altaffiltext{3}
{Jet Propulsion Laboratory, California Institute of Technology, Pasadena,
CA 91109}\
Doris Rosenbaum,\altaffilmark{1}\\
Randall J. Scalise,\altaffilmark{1}\ and
Paul Wentzler\altaffilmark{1}}

\begin{abstract}
We compute the temperature and IR signal of particles of radius $a$ and
albedo $\alpha$ at heliocentric distance $R$, taking into account the
emissivity effect, and give an interpolating formula for the result.  We
compare with analyses of COBE DIRBE data by others (including recent
detection of the cosmic IR background) for various values of heliocentric
distance, $R$, particle radius, $a$, and particle albedo, $\alpha$.  We then
apply these results to a recently-developed picture of the Kuiper belt as a
two-sector disk with a nearby, low-density sector ($40<R<50-90 AU$) and a
more distant sector with a higher density.  We consider the case in which
passage through a molecular cloud essentially cleans the Solar System of
dust.  We apply a simple model of dust production by comet collisions and
removal by the Poynting-Robertson effect to find limits on total and dust
masses in the near and far sectors as a function of time since such a
passage. Finally we compare Kuiper belt IR spectra for various parameter
values. Results of this work include: (1) numerical limits on \kb dust as a
function of ($R, a, \alpha$) on the basis of 4 alternative sets of
constraints including those following from recent discovery of the cosmic IR
background by \cite {ha98}; (2) application to the two-sector \kb model
finding mass limits and spectrum shape for different values of relevant
parameters including dependence on time elapsed since last passage through a
molecular cloud cleared the outer Solar System of dust; and (3) potential use 
of spectral information to determine time
since last passage of the Sun through a giant molecular cloud.
\end{abstract}
\section{Introduction}

        Two papers examining IR flux upper limits from COBE DIRBE and IRAS
(\cite{bk95}; \cite{st96a}) have put quite stringent limits on the total
amount of mass in the form of dust (particles less than about one centimeter) in
the Kuiper belt (KB) region inside $\sim50\,$AU.  Roughly, they find that the
dust mass, $M_D$, is bounded by $M_D<10^{-5}M_{\oplus}$.  In both cases,
however, the results are based on specific, albeit quite reasonable, models
for the distribution of dust as a function of particle radius, $a$, and
heliocentric distance, $R$, and an assumption as to albedo, $\alpha$.  The
objectives of this paper are: (i) to address KB IR radiation from an
arbitrary distribution; (ii) to obtain COBE DIRBE limits on dust mass, $M_D$,
as a function of $a, R, \alpha$; (iii) to apply these results to a case of
particular interest for a variety of parameter values; and (iv) to take into 
account the recent analysis of
COBE DIRBE data by \cite{ha98} in which bounds at 60 and 100 microns, and 
values at 140 and 240 microns, for the cosmic infrared background have been 
found (see Table 1).  We note that recent reviews of the
Kuiper belt include Stern (1996c) and Weissman and Levison (1997).

        The case of interest explored here is that in which the near portion
of the Kuiper belt ($40<R<[50-90]\,AU$) is significantly depopulated.  Such a
structure has been shown (\cite{st96b}) to make it possible to understand
formation of the large bodies observed (\cite{jl}; \cite{jlc}) in the face of
the large eccentricities observed, which would imply erosive rather than
growth-promoting collisions in the present day $30-50$ AU region.  The
explanation (\cite{st96c}) appears to be an originally more massive region of
low eccentricities to permit growth of larger bodies in a short accretion
time, with the region later suffering physical and dynamical erosion, and
depopulation due to dynamical interaction with Neptune (\cite{ledu};
\cite{how}; Malhotra 1995; \cite{dulb}; \cite{mtm}; Levison \al 1997).  In
this picture, only the component of the Kuiper belt sufficiently far past
these effects would retain the surface density distribution, $R^{-2}$ to
$R^{-3/2},$ of the original solar nebula.  The near belt (e.g., $30-70$ AU)
would be significantly depopulated.  A key implication of this result would
be different IR signals from each of the two sectors since, on the average,
the dust in the far sector would be cooler, dimmer, and in orbits of much
smaller eccentricity. Accretion in the early Kuiper belt is also studied by 
\cite{klu}.

The plan of the paper is as follows.  In Section 2 below, we give our results
on dust mass limits as a function of heliocentric distance, particle radius,
and particle albedo ($R,\,a,\,\alpha$) from COBE DIRBE IR signal limits.  In
Section 3 we apply these results to the mass limits for the two sector
distribution discussed above.  In Section 4 we discuss the variation of
spectra with model parameters.  Section 5 contains our concluding discussion.

\section{IR Dust Signal}\label{ird}

        The IR intensity signal in the wavelength band $\lambda$ to
$\lambda+d\lambda$ from $N$ particles, each of radius $a$ and temperature $T$
at heliocentric distance $R$, spread uniformly in an ecliptic band from
$-\theta$ to $+\theta$ is given by
\begin{equation}
\label{bbem1}
I(N,\theta,a,R,T,\lambda)d\lambda=(\pi a^2/R^2)[N/(4\pi\,sin\theta)]
\epsilon_{IR}\,I_{BB} d\lambda
\end{equation}
here $I_{BB}$ is the radiation emitted by a black body per unit
area-time-wavelength-solid angle:

\begin{equation}\label{bbem2}
I_{BB}(\lambda,\,T)=2hc^2/[\lambda^5(e^{hc/\lambda\,kT}-1)]
\end{equation}
the emissivity factor, $\epsilon_{IR}$, which gives the suppression of
radiation of wavelength larger than the particle is assumed to be:
\begin{equation}\label{bbem3}
\epsilon_{IR}(a,\lambda)=[1,(a/\lambda)^n]\ \ \ \  [\lambda\,<a,\lambda\,>a].
\end{equation}
We investigate below the implications of choosing n to be 1 and 2.  A useful
discussion of emissivity and also absorptivity factors is given by \cite{bp}.  
Because IR observations integrate their view radially (i.e., view the radial
distribution in projection on the sky), in applying results based on equation
(\ref{bbem1}) to specific models, one must integrate over the model dependent
distribution in radius $a$ and heliocentric distance $R$.  We will do this in
Section 3 below. The temperature for each particle can be found from its
values of $a, R$ and $\alpha$.  Equating solar power absorbed with power
emitted gives the standard result
\begin{equation}\label{teq1}
\pi\,a^2(1-\alpha)L_{\odot}/(4\pi\,R^2)=4\pi\,a^2\int_0^{\infty}\,I(\lambda,T)
\epsilon_{IR}(\lambda)d\lambda
\end{equation}
where $L_{\odot}$ is the luminosity of the Sun. We introduce a variable y by
\begin{equation}\label{teq6}
        T=hc/(kay).
\end{equation}
Equation (\ref{teq1}) can then be written in the form
\begin{equation}\label{teq2}
        z=(1-\alpha)L_{\odot}a^4/(32\pi^2hc^2R^2)=F(y)
\end{equation}
Below we invert equation (\ref{teq2}) to find y, and hence the temperature
which is needed to use equation (\ref{bbem1}) to find the IR signal from the
dust as a function of z.  In equation (\ref{teq2}) we have (for n=1)
\begin{equation}\label{teq3}
        F(y)=[yf_4(y)+f_5(y)]/y^5
\end{equation}
where the functions $f_k$ are given by
\begin{equation}\label{teq4}
        f_k(y)=\int_y^{\infty}\,x^{k-1}\,dx/[e^x-1].
\end{equation}
Inverting \Eq (\ref{teq2}) can be done analytically for $z$ near zero and
infinity, with the results
\begin{equation}\label{teq7}
        y=(f_4(0)/z)^{1/4}
\end{equation}
near infinity, and
\begin{equation}\label{teq8}
        y=(f_5(\infty)/z)^{1/5}
\end{equation}
near $z=0$.  In equations (\ref{teq7}) and (\ref{teq8}), the asymptotic forms
are given in terms of the ordinary gamma function, $\Gamma$, and
the Riemann zeta function, $\zeta$, by
\begin{equation}\label{teq9}
        f_k=\Gamma(k)\zeta(k).
\end{equation}
        We have evaluated the integrals in \Eq (\ref{teq2}) numerically and
found for n=1 that a simple quintic approximation to $ln\,y$ represents $y$
to better than one percent over the interval \footnote{For higher and lower z
equations (\ref{teq7} and \ref{teq8}) are preferred.} $15>ln\,z>-15$. The
approximation formula is
\begin{equation}\label{scalise}
y=e^{A_0}z^B
\end{equation}
where
\begin{equation}\label{beqas}
B=\sum_1^5\,A_n(ln\,z)^n
\end{equation}
with $A_0=0.4624;\ A_1=-0.2433;\ A_2=-0.001846;\ A_3=0.0001243;
\ A_4=2.799\times10^{-6}$; and $A_5=-3.066\times10^{-7}.$ A similar 
procedure can be used for other values of n in equation (\ref{bbem3}). 
\cite{bp} give a different expansion, but the temperature predictions are the 
same.

        With the results of equations (\ref{teq7}, \ref{teq8}, \ref{scalise})
for particle temperature, we can use \Eq (\ref{bbem1}) to find the total IR
signal from a given mass $M$ in the form of particles of radius $a$, density
$\rho$, and albedo $\alpha$ located at heliocentric distance $R$.  We insert
in \Eq (\ref{bbem1}) that
\begin{equation}\label{nfromm}
        N=M/(4\pi\rho\,a^3/3).
\end{equation}

We can then compare the results of \Eq (\ref{bbem1}) with COBE DIRBE data. In
Section 3 we will integrate over a distribution in R and $a$ in the model
considered.  Here we provide ``model-independent," unintegrated limits .  We
make comparisons with ``data" four times: first we choose, as the constraint
on the KB IR signal, the full COBE DIRBE signal (as given in Figure 2 of
\cite{bk95}); second, we choose the constraints that result from
(model-dependent) subtraction of the signal from the asteroid belt as cited
by Backman et al.  (credited to, e.g., Reach 1992, 1988).  Our third choice
is the COBE DIRBE two sigma upper limits for $\lambda=$60 and 100 microns and
values for 140 and 240 microns found by \cite{ha98}.  Finally we choose the
set of standard deviations from \cite{ha98}. Our treatment of the Hauser et
al. results is motivated by the fact that their procedure for subtracting the
Solar System contribution to the IR signal did not include explicitly fitting
and subtracting a Kuiper belt contribution, so that the implications of their
work for the KB are not clear.  Further discussion of their treatment of
Solar System source subtractions is given in \cite{dwk} and \cite{kel}. The
values in the four bands for each of the four sets of constraints are given
in Table 1.

In Figures 1 and 2 we display the results of using equation (\ref{bbem1}) to
find the maximum mass $M(a,R,\alpha)$ (in units of $M_{\oplus}$) of material
in the form of particles of radius $a$ with albedo $\alpha$ at heliocentric
distance $R$.  We use a particle density of $1\,$gm cm${}^{-3}$ and take the
KB density distribution as constant in density to $\pm$ 1/3 radians and zero
outside this region.  Our results can, of course, be scaled to other values
for these parameters.  Figure 1 gives results for $\alpha=0.05$, which would
be the case if the dust is either made mainly from the surface of comets or
made sufficiently long ago that it has suffered the same radiation darkening
as is believed responsible for the low albedos of comet nuclei
(e.g. \cite{jo}).  For comparison we also give results in Figure 2 for
$\alpha=0.5$, which might obtain if the dust were made of relatively pure
ices.

In Figure 1a we give the curves for all four constraint sets of Table 1 for
$a=0.1\,$cm.  These curves do not depend on the value of $n$ in equation (3)
because little radiation is emitted of wavelength larger than $0.1\, $cm.  In
Figure 1b we give the results for particle radius $a=0.001$ for all four
constraint sets and for $n=1$ and $n=2$.  Figure 2 gives the corresponding
curves for $\alpha = 0.5\,$cm.  Finally, Table 2 gives dust temperatures, for
$n=1$ and $n=2$ for selected values of $a$, $R$, and $\alpha$.

   Note that: (1) Figures 1 and 2 give the total mass that, if located exactly 
at heliocentric distance $R$ (and optically thin), would give an IR signal 
that would exhaust the 
constraint, not a distribution or a mass 
density; (2) the bound on the dust mass is linear in the assumed bound on the
signal whereas the dust mass itself and the IR signal from it, in models for
dust production such as those of Section 3 below, are proportional to the
square of the total mass of colliding comets that gives rise to the dust 
(as noted in \cite{bk95});  (3) 
our model for dust is unsophisticated compared to work addressing dust 
properties  in more detail such as \cite{lio}, \cite{ish}, and \cite{yamu}; 
(4) in the 
absence of the emissivity effect of equation (\ref{bbem3}), the mass limit 
M(\ara) scales linearly with $a$ (the ratio of volume to surface area).  The
emissivity effect can raise or lower mass limits according to the relative
values of long and short wavelength signal limits, since its effect is to
shift energy from long wavelength bands to short; and (5) the places in Figures
 1
and 2 at which there are discontinuities in slope are places at which there
is a change in which wavelength band is limiting.

        Some of us, in other work (\cite{adr97}) have looked at Kuiper belt
information that can be derived from Pioneer 10 which has been in the region
of the KB for over a decade.  A limit on the mass of the KB in the form of
objects of radius somewhat under a centimeter of roughly a tenth of an Earth
mass (averaged over the region of the KB out to about 65 AU) has been found
based on the fact that Pioneer encountered no object energetic enough to
penetrate its propellant tank.  That result is clearly compatible with the
results of this paper (i.e., that result is also an upper bound although one
less stringent than the IR bound of this paper).

\section{Two Sector Belt Model}

        We now apply the results above to the Kuiper belt model outlined in
the Introduction, in which there is an under populated near region and a
denser far region. We assume that the boundary is somewhere between 50 and 90
AU depending on such factors as the influence of Neptune on orbits and the
effects of velocity evolution in the disk. For this paper we attempt to
bracket the boundary by presenting results for it being at 50 AU and at 90
AU; where we need a representative value, we choose 70 AU.

        We insert the results of Section 2 for the IR signal as a function of
heliocentric distance $R$ and particle radius $a$ into a simple computer
model for dust production.  Our model for such production is similar to those
described previously (\cite{bk95}; \cite{st96a}), but extends them to take
into account dependence on time since passage through a GMC last cleansed the
Solar System of dust.  We assume that the region is populated, at first,
solely by planetesimals (comets) of $5\,$km radius, compute the collision
rate, assume that all collisions result in a dust distribution with an
$a^{-3.5}$ dependence, (in Section 4 we consider $a^{-3.0}$ as well) and that
each dust particle is removed at its appointed Poynting-Robertson (P-R) time.
More precisely, the model is as follows.

        The total collision ($dN_{col}(R)/dt$) rate is found from the
expression $N_1n_2\sigma v$ for collisions per unit time of $N_1$ particles
with a distribution $n_2$ per unit volume of others, with $\sigma$ the
collision cross section and $v$ the relative velocity. For us, $n_2$ is equal
to $n_C$, the number density of comets, and $N_1=n_C 4\pi R^2dR\sin\theta$.
For the collision of 5 km comets it is
\begin{equation}\label{collis}
d^2N_{col}(R)/dRdt=n_{C}^2(R)\pi\,a_C^2\,v_C (R)\,(4\pi\,R^2\,sin\theta)
\end{equation}
In equation (\ref{collis}), $n_C(R)$ is the number of comets per unit volume,
$a_C$ the 5 km radius, and $v_C$ the (average) relative velocity.  We take a
surface distribution, of total mass $M_T$ between $R_1$ and $R_2$, falling
like $R^{-2}$ to obtain
\begin{equation}\label{comdens}
        n_C(R)=(M_T/M_C)/[ln(R_2/R_1)4\pi\,sin\theta\,R^3]
\end{equation}
where $M_C$ is the mass of an individual comet (about $5\times 10^{17}$ gm
for ours, assuming unit density) and $\theta$ is the half-height extent of
the KB in declination.  We take the average relative velocity in equation
(\ref{collis}) to be given in terms of the average eccentricity $e$ by
\begin{equation}\label{vofr}
        v_C=\sqrt{2}v_K(R)\left[\frac{5}{4}<e>^2+<i>^2\right]^{1/2}
\end{equation}
where $v_K$ is the Kepler velocity, the $\sqrt{2}$ factor sums in quadrature
the velocity deviations from circular of the two colliding objects, and the
sum in square brackets is from Lissauer and Stewart (1993). Below, we make
the standard equilibrium approximation $<i>=<e>/2$.

In our model, the IR signal from the KB region between $R_1$ and
$R_2$, from particles of radii between $a_1$ and $a_2$ is given by
\begin{equation}\label{irsig}
I(\lambda)=\int_{R_1}^{R_2}\,dR\,\int_{a_1}^{a_2}\,da[d^2N_{col}(R)/dRdt]
[dN_D/da]\tau(R,a,t)I_1(a,\ R,\ \theta,\ \alpha).
\end{equation}
In equation (\ref{irsig}), $I_1$ is the wavelength distribution of equation
(\ref{bbem1}) for $N=1$ (i.e. for each particle). Here, $\tau$ is the shorter
of: (i) the time $t$ since last passage through a molecular cloud eliminated
Kuiper belt dust (\cite{st90}); and (b) the Poynting- Robertson time,
$t_{PR}$, for a particle of radius $a$ at heliocentric distance $R$.  (See
\cite{bn79} for a full, lucid discussion of this and related effects).  The
P-R time, $t_{PR}$, is given by

\begin{equation}\label{tpr}
        t_{PR}=4\pi\,c^2\rho\,a\,R^2/[3L_{\odot}(1-\alpha)]
\end{equation}
which may be evaluated approximately by
\begin{equation}\label{tpr2}
t_{PR}\cong 10^{10}yr\rho(gm\,cm^{-3})a(cm)(R(AU)/40)^2/(1-\alpha).
\end{equation}

Other effects that need, in principle, to be included are radiation pressure
and erosion by the ISM.  Radiation pressure is only important for $a$ on the
order of a few microns where the P-R time is already very short so little
error is made in neglecting it.  Stern (1990) has shown that most of the ISM
effect comes from occasional passage of the Sun through a giant molecular
cloud (GMC). We assume that the ISM effect can be approximated by periodic
removal of dust from the Solar System and compute dust production as a
function of time since the most recent such ``tidying."  There is an important 
question with regard to this assumption: GMC passage erodes about 10 
g\,cm$^{-2}$ from comets in the Oort cloud (\cite{st90}) and for ones in the 
KB.  For a KB of 0.1$M_{\oplus}$ in 5 km comets this implies creating a 
dust mass of about $10^{-5}M_{\oplus}$ -- of the order of the maximum dust 
mass compatible with IR signal bounds (Backman et al. 1995, Stern 1996).
 However the high relative velocity 
between the Solar System and the GMC (about 30\,km/s) means that the dust 
generated by the collisions will tend to have high enough velocity (roughly 
10\% of the 30\,km/s) to exit the Solar System (\cite{fuj}) so that much less 
bound dust will be generated (and that which is generated will be expelled in 
the next encounter). 

	We assume this Solar System bound dust is 
negligible.  This assumption is conservative in the sense that, if
some of this dust were captured into solar orbit, our upper limits
(below) on total KB mass would be made lower (more constraining). We take the
distribution $dN_D/da$ to be the one that is self replicating under
collisions (Dohnanyi 1969 and, more recently in more generality, \cite{tan}); 
in terms of mass it is given by $m^{-11/6}$, while
in terms of radius, we have
\begin{equation}\label{dnda}
dN_D/da=n_0a^{-7/2}.
\end{equation}
In Section 4, we briefly discuss the effect of varying from equation
(\ref{dnda}).  We assume density $\rho=1\,$gm/cm$^3$ for both comets and dust
and we use \Eq(\ref{dnda}) up to 5 km. Then, by normalizing equation 
(\ref{dnda}) to the total volume of the two comets, we have (with $a$ in cm)
\begin{equation}\label{nzero}
        n_0=a_c^{5/2}/2\approx2\times 10^{14}.
\end{equation}

        We have coded equations (\ref{collis})-(\ref{nzero}).  For each
set of initial parameters, we evaluated the IR signal in the four 
bands for a low value of total KB sector mass, and then increased the 
sector mass until the IR limit was exceeded in one of the four bands, 60, 
100, 140, 240 microns (for one of the four constraint sets of Table 1).  The 
results are given in Tables 3 and 4 and Figures 3 and 4. In the tables, the 
columns are: (1) ``case number;" (2) signal constraint applied (1-4 as in 
Table 1); (3) albedo $\alpha$; (4) $n$ in \Eq (\ref{bbem3}); (5) inner radius 
of sector; (6) outer radius of sector ; (7) time since dust 
began accumulating; (8) maximum sector mass in Earth masses; (9) total 
dust produced (one micron to 
0.5 centimeter) in units of $10^{-5}M_{\oplus}$; (10) dust remaining after 
Poynting-Robertson effect in the same units.  Table 3 gives results for 64 
sets of input parameters for the near sector of the Kuiper belt and Table 4 
for the far.  In each, the first 32 cases are for $n=1$ the last 32, $n=2$. 
We have taken two values for the time $\tau$ since the KB was last cleaned
of dust: the age of the Solar System and 200 million years.  The latter value
is about twice the minimum needed for comet nuclei radiation mantles to
darken their albedos to the $0.05$ average observed in the few cases for
which there are measurements (\cite{jo}). As noted above, we take 50 and 90
AU as extreme possibilities for the boundary between the near and far
sectors. As before, we take the KB comet number density to be constant in
declination to $\theta =\pm 1/3\,$radians for the near sector, For the far
sector, which did not suffer dynamical interaction with Neptune as did the
near, we use $<e>=\theta=0.01$ radians. 

	We turn first to Table 3 and the near sector.  Features of the 
calculations and results that might be noted include:

        --As expected, the IR signal rises as the square of the
total belt mass, $M_T$ (or the comet number density), since the collision
rate does.

        --As $<e>$ (or $\theta$) is increased, the relative velocity goes up
linearly, but the density goes down linearly so the total collision rate in 
equation (\ref{collis}) stays constant; the signal in equation (\ref{irsig}) 
then varies as $M_T^2/e$ because of the $\theta$ dependence of $I_1$. 
Thus the table can be extended approximately to other eccentricity 
values (within our simple model) by $M_T \propto (e)^{1/2}$.

        --We have good agreement with the results of Backman \al Case 15 in
Table 3 ($M_T=1.17$ for $\alpha=0.5, t=4.5\,$Gyr) corresponds to their
parameter and constraint choices but needs to be corrected for their choice
of $\rho=0.5$, $\theta=1/6$, and boundaries of 30-100 AU for them versus
40-90 AU for case 15. This gives a limit of about $M_{\oplus}/2\ [\cong
1.17(1/2)(1/\sqrt{2})(10/9)^{3/2}]$ which is their result ($M_{\oplus}/3)$
within better than a factor of 2.

        --The ratio of dust remaining (after P-R loses) to total produced,
$M_{DR}/M_D$, goes up with increasing $R^2$ because the Poynting-Robertson
time goes up as $R^2$. The masses $M_{DR}$ and $M_D$ each scale as $<e>$
since they vary as $M_T^2$.

        --There is a factor of about 30 between the largest near
sector upper mass limit, 7.0$\,M_{\oplus}$ (case 40), and the smallest, 0.22  
(case 29), but all values are above current best estimates of about 
0.1$\,M_{\oplus}$ for the near sector.

        --There is a factor of about 70 between the largest calculated value
of remaining dust, about $41\times10^{-5}M_{\oplus}$ (case 39), and the
smallest, about $0.56\times10^{-5}M_{\oplus}$ (case 26).

        --We have given results for $\alpha=0.05$ and $\alpha=0.5$ for
comparison although it is unlikely that the latter can obtain in reality.

	--Note the relative insensitivity to the emissivity index, n; going
from $n=1$ to $n=2$ raises the temperature of small grains significantly, but
it suppresses the rate of emission from them in the wavelength range above 
their size; these two effects tend to cancel.

        --We have chosen the upper limit for particle radius at $a_2=0.5\,cm,$
but one can understand the dependence of the IR signal on $a_2$ simply.  For
$n=n_0a^{-\beta},$ the IR signal goes as
$\int^{a_2}_{a_1}n_0a^{-\beta}a^2t_{PR}(a)da \sim a_2^{4-\beta}$ for
$t_{PR}(a_2)<t_0$ the present age of the Solar System.  The IR signal stops 
growing with $a_2$ when $t_{PR}(a_2)=t_0$. From equation (\ref{tpr}) we  
see that, for $\rho=1$, the signal stops growing with $a_2$ at $a_2=0.5\,$cm 
for  $\alpha\ll\,1$ (as noted by Backman et al. 1995).

	--Note that we can see, from Equations (\ref{collis}), (\ref{irsig}), 
(\ref{dnda}), and (\ref{nzero}), the effect of varying our assumption of 
$5\,km$ for the comet radius, $a_C$.  We have, by equation (\ref{nzero})
 $n_0 \sim a_C^{5/2}$ and 
$d^2N_{col}/dRdt \sim M^2/a_C^4$ which gives $I \sim 
M^2/a_C^{3/2}$ or $M_T \sim a_C^{3/4}$.

        --As emphasized by Backman (1995, 1998), there is a strong
dependence of the IR flux on the assumption made for the inner edge of
the dust distribution.  In our calculations, we assume that it is at
40 A.U. where Duncan, Levison and Budd (1995) find the approximate
boundary between stable and unstable orbits (Backman et al (1995) and
Stern (1996) assume a hard inner edge at the orbit of Neptune).  Were
it the case that the P-R effect operated unimpeded, then, in the
steady state, the signal from KB dust in the inner Solar System would
swamp the signal from KB itself.  Estimating the amount of dust that
survives into the inner Solar System is difficult.  Important effects
include: dynamical orbit stability as a function of grain mass;
collisions with interstellar dust (see Liou, Zook, and Dermott 1995);
and, particularly, sublimation of volatiles even out as far as 30 A.U.
Thus the results of this section on IR bounds on the KB mass should be
interpreted as quite conservative.  A substantial signal from KB dust
in the inner Solar System would mean much stronger bounds on the KB
mass and/or the time since last GMC encounter.

Table 4 gives the corresponding results for the far sector of the
KB.  We take the outer boundary of the far sector to be 120 AU and,
again, use $50$ and $90$ AU for the (inner) boundary with the near sector. 
One sees that the model limits on total sector mass are similar to 
those on the near sector.  The fact that they are not larger is a result of 
several factors including: $e$ and $\theta$ are assumed down by a factor of 30 
(cutting $M_T$ by a factor of $\sqrt{30}$ ); P-R dust losses fall quadratically 
with $R$; and the emissivity effect of equation (\ref{bbem3}) causes 
temperatures to fall less rapidly than $R^{-1/2}$ for small $a$.
It is very important to note that the dust production model may not apply to 
the far sector. This is because, for low values of $e$, one expects 
collisions to be adhesive rather than fragmenting in nature with little dust 
production. The definition of ``low,'' however, depends on the unknown (but
size-dependent strength of Kuiper materials.   As noted above, the limits for 
 $<e>$ high enough to result in 
fragmentation can be found from the results given by using the 
$<e>^{1/2}$ scaling law.  Thus, for example, if we take $100\,m/s$ as a rough
lower limit for strong materials, then for $<e>$ greater than $0.033$ 
at $R=100\,$AU the results of Table 4 (and Figure 3b and 4b below) scaled by 
$[<e>/0.01]^{1/2}$ would obtain. 

We turn now to consideration of the time dependence of the IR signal.  In 
Figure 3 we give the upper bounds on the near and far sector masses for the 
four constraint sets of Table 1.  Here we choose the sector boundary at 70 AU. 
Again we plot the results for the case of eccentricity $e=1/3$ for the near 
sector and $e=0.01$ (assuming collisions result in fragmentation) for the far 
sector. In computing the IR signal we have used, for albedo, $\alpha=.05 
$ at large times and $\alpha=0.5$ at small times with the break at $t_D=10^8$ 
years in order to see maximum possible time variation. However,
it is unlikely that even passage through a giant molecular cloud could lead
to an $\alpha$ value as high as 0.5. In computing the Poynting-Robertson
losses we used an interpolating function for $\alpha$ in equation(\ref{tpr})
\begin{equation}\label{alft}
\alpha(t)=0.5-0.45\,(1-e^{-t/t_D}).
\end{equation}

We see from Figure 3a that near sector, upper mass bounds decrease by about a
factor of 2 from short times ($10^7$ years) to $10^8$ years and about another
factor of 5 to long times ($4.5\times 10^{10}$ years.) The bounds on the far
sector, Figure 3b, fall somewhat more steeply.

In Figure 4 we plot total dust produced and dust remaining for the two
sectors as a function of time since last passage through a molecular cloud
for the case of constraint set 2.  One sees from the divergence of the
``total produced" and ``dust remaining" curves that, in time, the total mass
of dust removed by the P-R effect becomes considerable.  In Section 4 we will
see that P-R removal of small particles, which by the emissivity effect of
equation (\ref{bbem3}) have higher temperatures, results in a change over
time in the spectrum at long wavelengths. This is because the amount of high
T material stays constant while the amount of low T material grows.

A final effect of interest is potential destruction (Liou \al 1996) of KB dust 
grains of radii $4.5-25\,\mu$m by 26\,km/s ISM grains in the mass region 
$10^{-15}-10^{-12}\,$gm discovered by the space craft Ulysses 
(\cite{gru}). If the flux of such grains is relatively constant over time and 
position, removal of KB grains in the $4.5-25\,\mu$m range would decrease the 
IR signal significantly for the case t=t$_{PR}$.  For example, if it has been 
$10^8$ years since last GMC passage, this effect  could significantly 
decrease the signal from 80\,AU or more (assuming that $<e>$ is high enough to 
produce dust).
\section{Spectra}

In this section we explore a little, within the two sector model, the extent
to which spectral information in a Kuiper belt signal could provide
information on belt structure and properties or, contrariwise, on how the KB
spectrum varies with belt parameters.  First, Figure 5 shows the spectrum, of
each sector, for three different values of $\rho$ (0.5, 1.0, 2.0).  The
calculations are for n=1 and $\alpha$=0.05 and sector mass of
0.1$\,M_{\oplus}$.  Note that doubling $\rho$ halves the signal for small
wave lengths, but quarters it for long wavelengths. This is because short
wavelength radiation is emitted by small particles heated by the emissivity
effect of equation (\ref{bbem3}). Doubling $\rho$ for fixed sector mass
results in halving the number of comets and hence quartering the number of
collisions and the amount of dust. Since it also doubles the dust P-R
lifetime, the signal from small particles is only halved.  For large
wavelength, the signal is quartered because it comes largely from particles
with $t_{PR}$ already near, or greater than, $t_0$.

Figures 6 ($n=1$) and 7 ($n=2$) 
give spectra for the near and far sectors, for 0.2 and 4.5
Gyr since molecular cloud passage, and for two values of $\beta$ in the
particle distribution with radius ($n_0a^{-\beta}).$ We choose, for the two
values of $\beta,\ \beta=3.5$ as used in Section 3 and $\beta=3.0$.  this
latter value was used by Weissman and Levison (1997) to fit observations and
bounds for sizes over 1 km.  Thus it is somewhat speculative to apply it in
the sub-cm range.  Nevertheless it provides a measure of the variation in
bounds and in spectra with $\beta$ and it is of some special interest since
it sits on the dividing line between signal domination by large particles and
signal domination by small particles.  For $\beta$=3 the signal integral,
$\int^{a_2}_{a_1}\,da\,a^{-\beta}a^2$, for $a_2>a_{PR}$
($t_{PR}(a_{PR})=t_0)$ varies logarithmically with $a_2$.  We use for $a_2$, in
this case, the same $0.5cm$ as for $\beta=3.5$ for comparison purposes, rather
than the 10 km above which the Weissman-Levison fit has $\beta$ changed to 4.5
(the correction would be an overall factor of about 10).

In Figures 6 and 7, we have chosen $70\,$AU for the boundary between near and
far sectors, $e=1/3$ and 0.01 respectively for the two sectors, $\rho=1$, a 
mass of 0.1$\,M_{\oplus}$ for both sectors in all cases, and 
albedo $\alpha=0.05$.  Figure 6 has emissivity index  $n=1$ in equation
(\ref{bbem3}); Figure 7, $n=2$. The first observation to be made with regard to
 the results is
the striking variation among the spectra according to time since last passage
through a GMC.  The 0.2 Gyr and 4.5 Gyr intensity curves agree for $\lambda
<20$ microns, but differ by over an order of magnitude for $\lambda >100$
microns.  Therefore the IR Kuiper belt signal spectrum would appear to hold
significant promise for determining the time since such passage.

We also see that the $\beta=3.0$ spectra, for short wavelength, fall below the 
$\beta=3.5$ spectra
by about roughly the expected value $(10\,$km/$1\,$cm)$^{1/2}\sim10^3$ where
1 cm is the effective upper limit for the $a$ integral for $\beta=1$ and 10
km is the upper limit for $\beta=3$. Note that the factor of 33 in the ratio of
the values chosen for $<e>$ for the two sectors roughly cancels the effect of
the greater distance to the far sector.  One might also recall from the
discussion of Section 3 that the intensity curves scale with $M_T, <e>,$
and $a_C$ as $M_T^2/(<e>a_C^{3/2}$

	In Figure 8, we compare spectra for times
of $10^7,\,2\times10^8,$ and $4.5\times10^9$ years for near and far sectors
(using: $\beta=3.5,\,n=1;\,\alpha=0.04;\,M=0.1\,M_{\oplus};$ and
$<e>=0.333,\,0.01$ for near and far sectors respectively).  Again the ratio of
the values of $<e>$ chosen makes the intensities for the two sectors similar to
within an order of magnitude.  In both cases we see that for wavelengths around
30 microns, the ratio of the intensities for the two extreme times is about a
factor of 4 while that for $\lambda>200\,\mu m$ is over 100.  Future 
measurements
of the KB IR spectrum should provide, in addition to much compositional
information, information on the length of time since the belt was last cleaned
of dust.

As in Section 3, we have assumed that the inner
edge of the KB dust distribution is at 40 A.U.  If the P-R effect, in spite
of the dust destruction effects such as sublimation cited there, were to 
result in substantial amounts of KB dust pentrating to the inner Solar System 
there would be corresponding modifications in the spectra as viewed from the 
Earth.

\section{Conclusion}

        It is tempting, in thinking about IR constraints on the Kuiper belt,
to work to the result (\cite{bk95}; \cite{st96a}) of about
$10^{-5}M_{\oplus}$ of dust of radii between $1 \mu\,$m and 5 mm.  However
the results of Sections 2 through 4 show that rather wide ranges of values
for total Kuiper belt mass (factor of 30) and for belt dust (factor of 70) are 
possible. Tables 3 and 4 and
Figures 1-3 show that there can be major differences in this mass limit
according to assumptions made about albedo, distribution in particle size,
and heliocentric distance, subtraction of IR signal from non-belt sources,
emissivity effect index, belt inclination, and time since last passage
through a molecular cloud.  As discussed above, our results are in the
approximation that the effect of the ISM can be approximated (Stern 1990) by
removal of Solar System dust upon passage through a giant molecular cloud. 
Survival of dust from such an encounter and/or subsequent ISM destruction or
creation of KB dust could change our conclusions, but we do not
believe these effects will make major changes for the reasons given above.
As noted in Section 3, our mass limits are
quite conservative in that we do not include the IR signal from dust particles
that might, under the P-R effect, penetrate to the inner Solar System in spite
of such dust destruction effects as sublimation of volatiles.

If a surface mass density of the form $\Sigma (R)=\Sigma_0 R^{-2}$ is
normalized to the mass of solids in the outer planets between 5 and 35 AU
(about $48\,M_{\oplus})$ we have $\Sigma_0\sim3.9M_{\oplus}\,$AU${}^{-2}$. 
Such a distribution would give about $13M_{\oplus}$ between 40 and 70 AU and 
the same amount between 70 and 120 ( and
a little over half that much between 90 and 120). One can see from  
Figure 3  that the
IR limits make it unlikely that the near sector of the Kuiper belt can
have the full $\Sigma_0 R^{-2}$ surface density distribution of the inner
planets. If the far sector is dynamically cool enough that collisions are 
adhesive rather than fragmenting, there is little dust and no real IR limit on 
sector mass.  

The spectra studied in Section 4 tend to indicate that future spectral
information could help distinguish Kuiper belt parameters. In particular, one
result of the Poynting-Robertson effect is that such spectrum
characterizations as the ratio of long wavelength intensity to short
wavelength intensity could be a good measure of time since last passage of
the Sun through a giant molecular cloud.

\acknowledgments
It is a pleasure to thank Yiannis Kontoyiannis for advice on making figures
and Eli Dwek for crucial communications on the cosmic IR background. 
A wonderful referee's report by Professor D. E. Backman was of great 
assistance.
The work of VLT was supported in part by DOE grant DE-FG03-95ER40908.


\newpage
\figcaption{(a) Model-independent mass limits (in units of $M_{\oplus})$
for particles of radius $a=0.1\,$cm and albedo $\alpha=0.05$ as a function of
heliocentric distance $R$ for the 4 sets of constraints of Table 1.  The
curves for the case $ n=1$ in equation (\ref{bbem3}) are essentially
identical to those for $n=2$ as explained in the text. The digit by each
curve is the constraint set number. (b) Limits (in units of $M_{\oplus}$) for
particles of radius $a=0.001\,$cm and albedo $\alpha=0.05$ as a function of
heliocentric distances for the 4 sets of sector constraints.  Curves with $
n=1$ in equation (\ref{bbem3}) are not identical to ones with $ n=2$ as in
Figure 1a because the equilibrium temperatures for large bodies at the
heliocentric distances involved produce significant IR with
$\lambda>a=0.001$. The pair of digits by each curve gives constraint number
and $n$.}
\figcaption{(a) Limits for $a=0.1\,$cm with $\alpha=0.5$, as explained in
the caption for Figure 1a. (b) Limits for $a=0.001\,$cm with $\alpha=0.5$ as
explained in the caption for Figure 1b.}
\figcaption{(a) Upper limit on near sector mass as a function of time since
last passage through a GMC for the 4 different constraint sets of Table 1 (in
 units of Earth masses). Note $<e>=0.333$.  (b) Upper limit on far sector
 mass (as in Figure 3a). Note $<e>=0.01.$}
\figcaption{Dust produced and dust remaining (after P-R effect) as a function
of time since last passage through a GMC (in units of $M_{\oplus}$) for both
near and far sectors assuming saturation of the mass bound for constraint set
2 of Table 1. Here $<e>$ is 0.333 for near sector and 0.01 for far.}
\figcaption{Spectrum $I(\lambda)$ variation with density $\rho$ (of both
comets and dust) for near and far sectors.  Note, as explained in text,
$I\sim\rho^{-1}$ for short $\lambda$ but $\rho^{-2}$ for long.}
\figcaption{Spectrum variation with (i) sector [S=(N,F)=(near, far)], (ii)
time since last GMC passage [t=(R,D)=(recent,
distant)=($2\times10^8,\,4.5\times10^9)\,$yr], and (iii) number density
distribution index ($a^{-\beta},\beta=3.0,3.5)$ for albedo $\alpha=0.05$ and
$n=1$ in \Eq 3. Curves are labeled by (S,t,$\beta$).  For reference, in
descending order at $\lambda=0.1$~cm we have: (F,D,3.5), (N,D,3.5), (F,D,3.0),
(F,R,3.5), (N,R,3.5), (N,D,3.0), (F,R,3.0), (N,R,3.0).}
\figcaption{Spectrum variation as in Figure 6 for  $n=2$. Note that the values
at $\lambda=0.1$~cm are just as in Figure 6 (as might be expected since
essentially all long wavelength radiation comes from $a>\lambda$ which is 
unaffected by the emissivity effect).}
\figcaption{Spectrum variation with (i) time since last GMC passage
[t=$10^7,\,2\times10^8,$ and $4.5\times10^9\,$yr ] and (ii) sector [S=(N,F)]
for $\beta=3.5,\alpha=0.05,$ and $n=1.$}
\begin{deluxetable}{cccccccc}
\footnotesize
\tablenum{1}
\tablecaption{COBE DIRBE Upper Limits on the IR Signal From the Kuiper Belt
in MJy\,sr$^{-1}$.}
\tablewidth{13.50cm}
\tablehead{\colhead{}&\colhead{}&\colhead{$\lambda(cm)$}&\colhead{$I_{\nu }$
\tablenotemark{a}}&\colhead{$I_{\nu }$\tablenotemark{b}}&\colhead{$I_{\nu}$
\tablenotemark{c}}&\colhead{$I_{\nu}$\tablenotemark{d}}&\colhead{}}
\startdata
&  & 0.006  &                 16.0 &                   0.3&1.50&0.75&  \nl
&  & 0.010  &                   8.0   &                  1.0&1.27&0.63&\nl
&  &0.014  &                  5.2   &                 2.5&1.17&0.32&  \nl
&  &0.024 &                   3.1  &                  2.0&1.12&0.24& \nl
\enddata
\tablenotetext{a}{Constraints from data at $-1/3\,$radians 
declination.}\tablenotetext{b}{Constraints subtracting the asteroid
belt contribution (\cite{bk95}).}\tablenotetext{c}{Constraints from
\cite{ha98}, CIB two -sigma limits ($\lambda$=0.06 and 0.01) and values
($\lambda$=0.014 and 0.024).}\tablenotetext{d}{Constraints from \cite{ha98},
CIB one sigma errors.}
\end{deluxetable}

\begin{deluxetable}{clccccc}
\footnotesize
\tablenum{2}
\tablecaption{Dust Temperature As A Function of $R,a,\alpha$ and
n.\label{tbtemp}}
\tablewidth{13.50cm}
\tablehead{\colhead{R}&\colhead{}&\colhead{T }&\colhead{T}&\colhead{T}&
\colhead{T}&\colhead{ T}\nl
\colhead{(AU)}&\colhead{$\ \ \ \ a$=}&\colhead{(1 cm)}& \colhead{(0.1cm)}&
\colhead{(0.01cm)}&\colhead{($10^{-3}$cm)}&\colhead{($10^{-4}$cm)}}
\startdata
n=1&$\alpha$=0.05&&&&&\nl
40&&   43.9&   43.9&   45.7&   67.5&  106.9\nl
60&&   35.9&   35.9&   38.0&   57.4&   90.9\nl
80&&   31.1&   31.1&   33.4&   51.1&   81.0\nl
100&& 27.8&   27.8&   30.3&   46.8&   74.1\nl
120&& 25.4&   25.4&   28.0&   43.5&   68.9\nl
&&&&&&\nl
n=1&$\alpha$=0.5&&&&&\nl
40&&   37.4&   37.4&   39.4&   59.3&   94.1\nl
60&&   30.6&   30.6&   32.9&   50.5&   80.0\nl
80&&   26.5&   26.5&   29.0&   45.0&   71.3\nl
100&& 23.7&   23.7&   26.4&   41.1&   65.2\nl
120&& 21.6&   21.6&   24.4&   38.2&   60.6\nl
&&&&&&\nl
n=2&$\alpha$=0.05&&&&&\nl
40&&  43.9&   43.9&   46.6&   86.2&  185.7\nl
60&&  35.9&   35.9&   39.1&   75.3&  162.2\nl
80&&  31.1&   31.1&   34.6&   68.4&  147.4\nl
100&&27.8&   27.8&   31.6&   63.5&  136.8\nl
120&&25.4&   25.4&   29.4&   59.8&  128.8\tablebreak
n=2&$\alpha$=0.5&&&&&\nl
40&&  37.4&   37.4&   40.5&   77.5&  166.9\nl
60&&  30.6&   30.6&   34.2&   67.7&  145.8\nl
80&&  26.5&   26.5&   30.4&   61.5&  132.4\nl
100&&23.7&   23.7&   27.8&   57.1&  123.0\nl
120&&21.6&   21.6&   26.0&   53.7&  115.7\nl
\enddata
\end{deluxetable}

\begin{deluxetable}{lcrcrrrrrr}
\footnotesize
\scriptsize
\tablenum{3}
\tablecaption{Total and Dust Masses for Near Sector.\label{tbnear}}
\tablewidth{14.5cm}
\tablehead{\colhead{Case}&\colhead{Type\tablenotemark{a}}&\colhead{$\alpha$}&
\colhead{n\tablenotemark{b}}&\colhead{$R_1$}&\colhead{$R_2$}&\colhead{t}&
\colhead{$M_{belt}$ }&\colhead{$M_D$}&\colhead{$M_{DR}$}\nl
\colhead{}&\colhead{}&\colhead{}&\colhead{}&\colhead{}&\colhead{}&
\colhead{(Gyr)}&\colhead{($M_{\oplus}$)}&\colhead{$(10^{-5}M_{\oplus})$}&
\colhead{$(10^{-5}M_{\oplus})$}}
\startdata

1&     1&   0.05&   1&  40.0&     50.0&  4.5&     0.86&    55.43&    26.37\nl
2&     1&   0.05&   1 & 40.0&     50.0&  0.2&     1.75&    10.18&     9.17\nl
3&     1&   0.05&   1&  40.0&     90.0&  4.5&     2.59&    66.90&    35.81\nl
4&     1&   0.05&   1&  40.0&     90.0&  0.2&     5.52&    13.46&    12.29\nl
5&     1&   0.50&   1&  40.0&     50.0&  4.5&     0.80&    47.28&    29.47\nl
6&     1&   0.50&   1&  40.0&     50.0&  0.2&     1.95&    12.59&    11.72\nl
7&     1&   0.50&   1&  40.0&     90.0&  4.5&     2.44&    58.94&    39.29\nl
8&     1&   0.50&   1&  40.0&     90.0&  0.2&     6.18&    16.84&    15.83\nl
9&     2&   0.05&   1&  40.0&     50.0&  4.5&     0.30&     6.50&     3.09\nl
10&   2&   0.05&   1&  40.0&     50.0&  0.2&     0.50&     0.81&     0.73\nl
11&   2&   0.05&   1&  40.0&     90.0&  4.5&     0.95&     8.95&     4.79\nl
12&   2&   0.05&   1&  40.0&     90.0&  0.2&     1.59&     1.11&     1.02\nl
13&   2&   0.50&   1&  40.0&     50.0&  4.5&     0.36&     9.73&     6.07\nl
14&   2&   0.50&   1&  40.0&     50.0&  0.2&     0.60&     1.18&     1.10\nl
15&   2&   0.50&   1&  40.0&     90.0&  4.5&     1.17&    13.66&     9.11\nl
16&   2&   0.50&   1& 40.0&     90.0&  0.2&     1.91&     1.62&     1.52\nl
17&   3&   0.05&   1&  40.0&     50.0&  4.5&     0.42&    13.12&     6.24\nl
18&   3&   0.05&   1&  40.0&     50.0&  0.2&     0.82&     2.21&     1.99\nl
19&   3&   0.05&   1&  40.0&     90.0&  4.5&     1.28&    16.38&     8.77\nl
20&   3&   0.05&   1&  40.0&     90.0&  0.2&     2.59&     2.95&     2.70\nl
21&   3&   0.50&   1&  40.0&     50.0&  4.5&     0.41&    12.69&     7.91\nl
22&   3&   0.50&   1&  40.0&     50.0&  0.2&     0.92&     2.83&     2.64
\tablebreak
23&   3&   0.50&   1&  40.0&     90.0&  4.5&     1.28&    16.34&    10.89\nl
24&   3&   0.50&   1&  40.0&     90.0&  0.2&     2.93&     3.79&     3.56\nl
25&   4&   0.05&   1&  40.0&     50.0&  4.5&     0.22&     3.58&     1.70\nl
26&   4&   0.05&   1&  40.0&     50.0&  0.2&     0.43&     0.63&     0.56\nl
27&   4&   0.05&   1&  40.0&     90.0&  4.5&     0.67&     4.48&     2.40\nl
28&   4&   0.05&   1&  40.0 &    90.0&  0.2&     1.37&     0.83&     0.76\nl
29&   4&   0.50&   1&  40.0&     50.0&  4.5&     0.22&     3.49&     2.17\nl
30&   4&   0.50&   1&  40.0&     50.0&  0.2&     0.48&     0.77&     0.72\nl
31&   4&   0.50&   1&  40.0&     90.0&  4.5&     0.67&     4.46&     2.98\nl
32&   4&   0.50&   1&  40.0&     90.0&  0.2&     1.53&     1.04&     0.97\nl
33&   1&   0.05&   2&  40.0&     50.0&  4.5&     0.88&    57.71&    27.46\nl
34&   1&   0.05&   2&  40.0&     50.0&  0.2&     1.88&    11.73&    10.56\nl
35&   1&   0.05&   2&  40.0&     90.0&  4.5&     2.65&    69.88&    37.40\nl
36&   1&   0.05&   2&  40.0&     90.0&  0.2&     5.99&    15.81&    14.43\nl
37&   1&   0.50&   2&  40.0&     50.0&  4.5&     0.82&    49.75&    31.01\nl
38&   1&   0.50&   2&  40.0&     50.0&  0.2&     2.19&    15.91&    14.81\nl
39&   1&   0.50&   2&  40.0&     90.0&  4.5&     2.50&    62.30&    41.53\nl
40&   1&   0.50&   2&  40.0&     90.0&  0.2&     7.02&    21.77&    20.47\nl
41&   2&   0.05&   2&  40.0&     50.0&  4.5&     0.30&     6.70&     3.19\nl
42&   2&   0.05&   2&  40.0&     50.0&  0.2&     0.51&     0.87&     0.79\nl
43&   2&   0.05&   2&  40.0&     90.0&  4.5&     0.96&     9.16&     4.90\nl
44&   2&   0.05&   2&  40.0&     90.0&  0.2&     1.65&     1.20&     1.09
\tablebreak
45&   2&   0.50&   2&  40.0&     50.0&  4.5&     0.37&     9.93&     6.19\nl
46&   2&   0.50&   2&  40.0&     50.0&  0.2&     0.62&     1.27&     1.18\nl
47&   2&   0.50&   2&  40.0&     90.0&  4.5&     1.18&    13.87&     9.24\nl
48&   2&   0.50&   2&  40.0&     90.0&  0.2&     1.98&     1.73&     1.63\nl
49&   3&   0.05&   2&  40.0&     50.0&  4.5&     0.43&    13.60&     6.47\nl
50&   3&   0.05&   2&  40.0&     50.0&  0.2&     0.87&     2.48&     2.23\nl
51&   3&   0.05&   2&  40.0&     90.0&  4.5&     1.31&    16.98&     9.09\nl
52&   3&   0.05&   2&  40.0&     90.0&  0.2&     2.76&     3.37&     3.07\nl
53&   3&   0.50&   2&  40.0&     50.0&  4.5&     0.42&    13.23&     8.25\nl
54&   3&   0.50&   2&  40.0&     50.0&  0.2&     1.02&     3.48&     3.24\nl
55&   3&   0.50&   2&  40.0&     90.0&  4.5&     1.31&    17.12&    11.41\nl
56&   3&   0.50&   2&  40.0&     90.0&  0.2&     3.30&     4.79&     4.51\nl
57&   4&   0.05&   2&  40.0&     50.0&  4.5&     0.22&     3.73&     1.77\nl
58&   4&   0.05&   2&  40.0&     50.0&  0.2&     0.47&     0.72&     0.65\nl
59&   4&   0.05&   2&  40.0&     90.0&  4.5&     0.68&     4.64&     2.48\nl
60&   4&   0.05&   2&  40.0&     90.0&  0.2&     1.49&     0.97&     0.89\nl
61&   4&   0.50&   2&  40.0&     50.0&  4.5&     0.22&     3.63&     2.26\nl
62&   4&   0.50&   2&  40.0&     50.0&  0.2&     0.54&     0.98&     0.91\nl
63&   4&   0.50&   2&  40.0&     90.0&  4.5&     0.69&     4.68&     3.12\nl
64&   4&   0.50&   2&  40.0&     90.0&  0.2&     1.74&     1.34&     1.26\nl

\enddata
\tablenotetext{a}{Table 1 constraint number.}
\tablenotetext{b}{See equation (3).}
\end{deluxetable}

\begin{deluxetable}{ccrcrrrrrr}
\footnotesize
\tablenum{4}
\tablecaption{Total and Dust Masses for Far Sector.}
\tablewidth{14.5cm}
\tablehead{\colhead{Case}&\colhead{Type\tablenotemark{a}}&\colhead{$\alpha$}&
\colhead{n\tablenotemark{b}}&\colhead{$R_1$}&\colhead{$R_2$}&\colhead{t}&
\colhead{$M_{belt}$ }&\colhead{$M_D$}&\colhead{$M_{DR}$ }\nl
\colhead{}&\colhead{}&\colhead{}&\colhead{}&\colhead{}&\colhead{}&
\colhead{(Gyr)}&\colhead{($M_{\oplus}$)}&\colhead{$(10^{-5}M_{\oplus})$}&
\colhead{$(10^{-5}M_{\oplus})$}}
\startdata
 1&   1&   0.05&   1&  50.0&    120.0&  4.5&     0.87&     2.92&     1.85\nl
 2&   1&   0.05&   1&  50.0&    120.0&  0.2&     2.09&     0.76&     0.71\nl
 3&   1&   0.05&   1&  90.0&    120.0&  4.5&     1.57&     7.45&     5.81\nl
 4&   1&   0.05&   1&  90.0&    120.0&  0.2&     4.75&     3.03&     2.93\nl
 5&   1&   0.50&   1&  50.0&    120.0&  4.5&     0.85&     2.78&     2.05\nl
 6&   1&   0.50&   1&  50.0&    120.0&  0.2&     2.38&     0.98&     0.94\nl
 7&   1&   0.50&   1&  90.0&    120.0&  4.5&     1.65&     8.22&     6.94\nl
 8&   1&   0.50&   1&  90.0&    120.0&  0.2&     5.38&     3.88&     3.80\nl
 9&   2&   0.05&   1&  50.0&    120.0&  4.5&     0.38&     0.57&     0.36\nl
10&   2&   0.05&   1&  50.0&    120.0&  0.2&     0.64&     0.07&     0.07\nl
11&   2&   0.05&   1&  90.0&    120.0&  4.5&     1.00&     3.04&     2.37\nl
12&   2&   0.05&   1&  90.0&    120.0&  0.2&     1.59&     0.34&     0.33\nl
13&   2&   0.50&   1&  50.0&    120.0&  4.5&     0.49&     0.93&     0.69\nl
14&   2&   0.50&   1&  50.0&    120.0&  0.2&     0.77&     0.10&     0.10\nl
15&   2&   0.50&   1&  90.0&    120.0&  4.5&     1.21&     4.40&     3.71\nl
16&   2&   0.50&   1&  90.0&    120.0&  0.2&     1.92&     0.49&     0.48\nl
17&   3&   0.05&   1&  50.0&    120.0&  4.5&     0.45&     0.78&     0.49\nl
18&   3&   0.05&   1&  50.0&    120.0&  0.2&     0.99&     0.17&     0.16\nl
19&   3&   0.05&   1&  90.0&    120.0&  4.5&     0.91&     2.50&     1.95\nl
20&   3&   0.05&   1&  90.0&    120.0&  0.2&     2.30&     0.71&     0.68\nl
21&   3&   0.50&   1&  50.0&    120.0&  4.5&     0.47&     0.85&     0.63\nl
22&   3&   0.50&   1&  50.0&    120.0&  0.2&     1.13&     0.22&     0.21
\tablebreak
23&   3&   0.50&   1&  90.0&    120.0&  4.5&     0.99&     2.97&     2.51\nl
24&   3&   0.50&   1&  90.0&    120.0&  0.2&     2.68&     0.97&     0.94\nl
25&   4&   0.05&   1&  50.0&    120.0&  4.5&     0.23&     0.21&     0.13\nl
26&   4&   0.05&   1&  50.0&    120.0&  0.2&     0.52&     0.05&     0.04\nl
27&   4&   0.05&   1&  90.0&    120.0&  4.5&     0.44&     0.58&     0.45\nl
28&   4&   0.05&   1&  90.0&    120.0&  0.2&     1.20&     0.19&     0.19\nl
29&   4&   0.50&   1&  50.0&    120.0&  4.5&     0.24&     0.22&     0.16\nl
30&   4&   0.50&   1&  50.0&    120.0&  0.2&     0.59&     0.06&     0.06\nl
31&   4&   0.50&   1&  90.0&    120.0&  4.5&     0.46&     0.64&     0.54\nl
32&   4&   0.50&   1&  90.0&    120.0&  0.2&     1.40&     0.26&     0.26\nl
33&   1&   0.05&   2&  50.0&    120.0&  4.5&     0.89&     3.08&     1.95\nl
34&   1&   0.05&   2&  50.0&    120.0&  0.2&     2.34&     0.95&     0.88\nl
35&   1&   0.05&   2&  90.0&    120.0&  4.5&     1.64&     8.09&     6.31\nl
36&   1&   0.05&   2&  90.0&    120.0&  0.2&     5.96&     4.76&     4.59\nl
37&   1&   0.50&   2&  50.0&    120.0&  4.5&     0.88&     2.98&     2.20\nl
38&   1&   0.50&   2&  50.0&    120.0&  0.2&     2.82&     1.38&     1.32\nl
39&   1&   0.50&   2&  90.0&    120.0&  4.5&     1.76&     9.32&     7.87\nl
40&   1&   0.50&   2&  90.0&    120.0&  0.2&     7.12&     6.80&     6.65\nl
41&   2&   0.05&   2&  50.0&    120.0&  4.5&     0.39&     0.58&     0.37\nl
42&   2&   0.05&   2&  50.0&    120.0&  0.2&     0.66&     0.07&     0.07\nl
43&   2&   0.05&   2&  90.0&    120.0&  4.5&     1.03&     3.21&     2.50\nl
44&   2&   0.05&   2&  90.0&    120.0&  0.2&     1.65&     0.36&     0.35
\tablebreak
45&   2&   0.50&   2&  50.0&    120.0&  4.5&     0.49&     0.95&     0.70\nl
46&   2&   0.50&   2&  50.0&    120.0&  0.2&     0.79&     0.11&     0.10\nl
47&   2&   0.50&   2&  90.0&    120.0&  4.5&     1.24&     4.68&     3.95\nl
48&   2&   0.50&   2&  90.0&    120.0&  0.2&     1.97&     0.52&     0.51\nl
49&   3&   0.05&   2&  50.0&    120.0&  4.5&     0.46&     0.82&     0.52\nl
50&   3&   0.05&   2&  50.0&    120.0&  0.2&     1.09&     0.21&     0.19\nl
51&   3&   0.05&   2&  90.0&    120.0&  4.5&     0.94&     2.68&     2.09\nl
52&   3&   0.05&   2&  90.0&    120.0&  0.2&     2.84&     1.08&     1.04\nl
53&   3&   0.50&   2&  50.0&    120.0&  4.5&     0.48&     0.90&     0.66\nl
54&   3&   0.50&   2&  50.0&    120.0&  0.2&     1.34&     0.31&     0.30\nl
55&   3&   0.50&   2&  90.0&    120.0&  4.5&     1.06&     3.37&     2.84\nl
56&   3&   0.50&   2&  90.0&    120.0&  0.2&     3.57&     1.71&     1.67\nl
57&   4&   0.05&   2&  50.0&    120.0&  4.5&     0.24&     0.22&     0.14\nl
58&   4&   0.05&   2&  50.0&    120.0&  0.2&     0.58&     0.06&     0.05\nl
59&   4&   0.05&   2&  90.0&    120.0&  4.5&     0.46&     0.63&     0.49\nl
60&   4&   0.05&   2&  90.0&    120.0&  0.2&     1.49&     0.30&     0.29\nl
61&   4&   0.50&   2&  50.0&    120.0&  4.5&     0.24&     0.23&     0.17\nl
62&   4&   0.50&   2&  50.0&    120.0&  0.2&     0.70&     0.08&     0.08\nl
63&   4&   0.50&   2&  90.0&    120.0&  4.5&     0.49&     0.72&     0.61\nl
64&   4&   0.50&   2&  90.0&    120.0&  0.2&     1.87&     0.47&     0.46\nl

\enddata
\tablenotetext{a}{Table 1 constraint number.}
\tablenotetext{b}{See equation (3).}
\end{deluxetable}
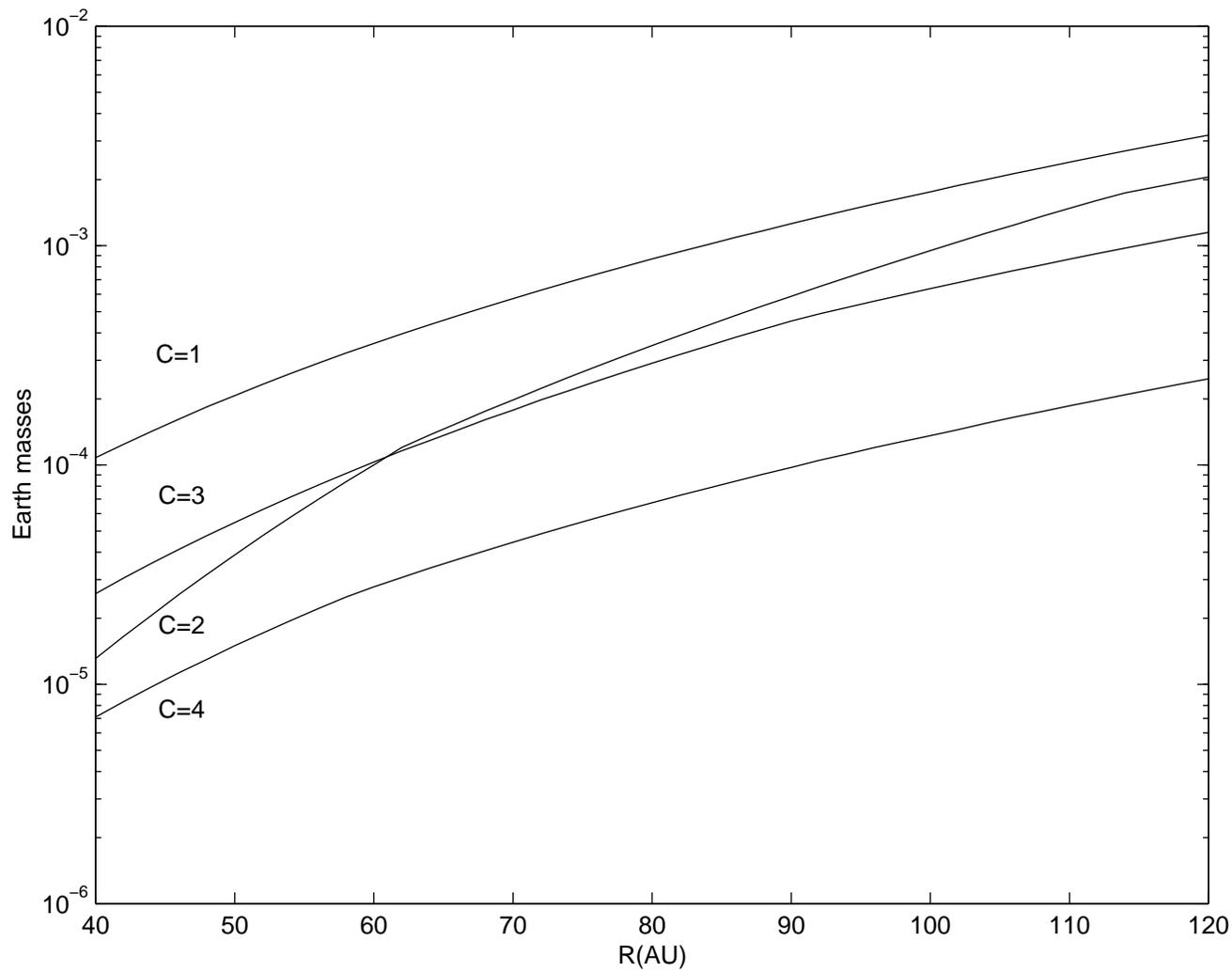
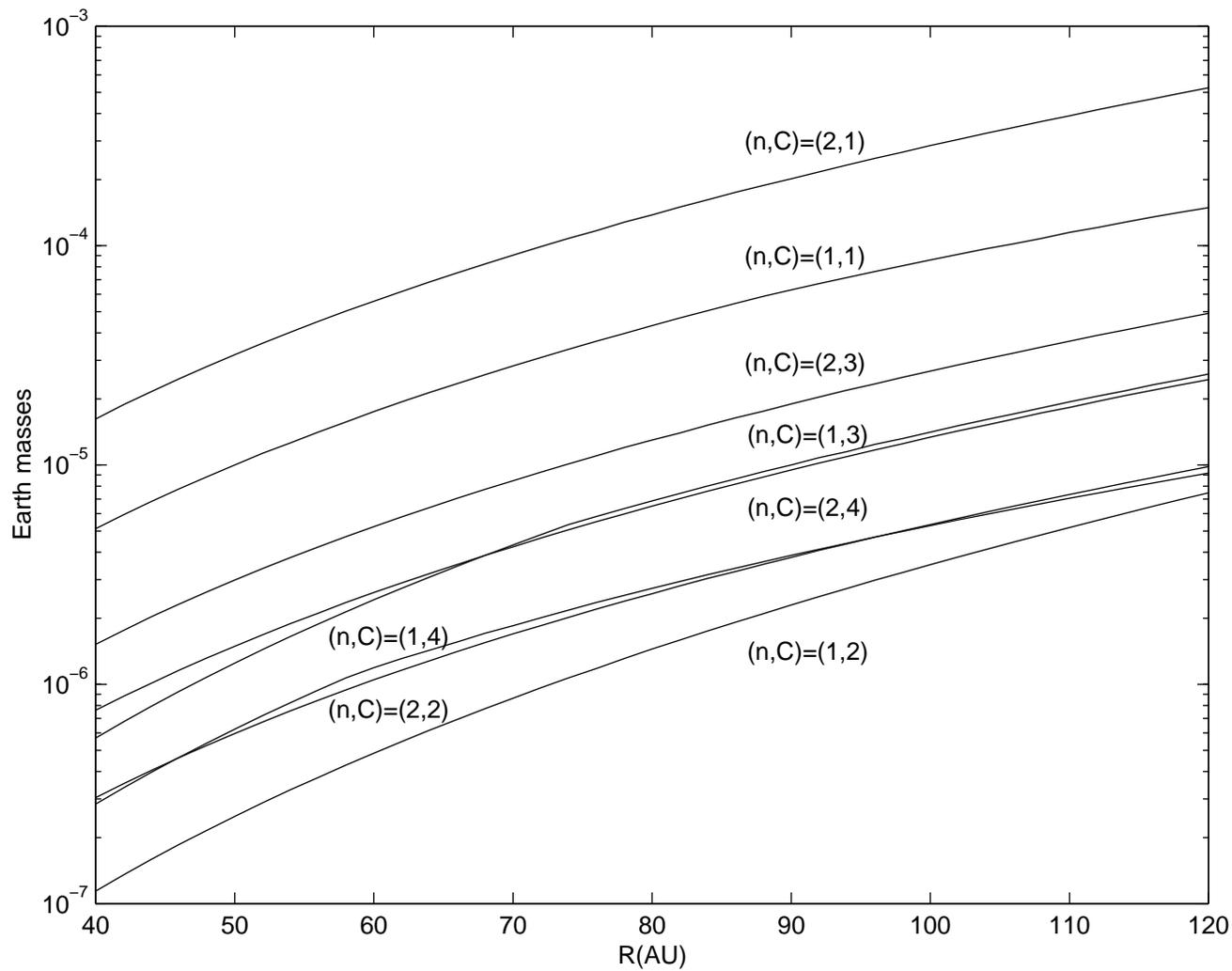
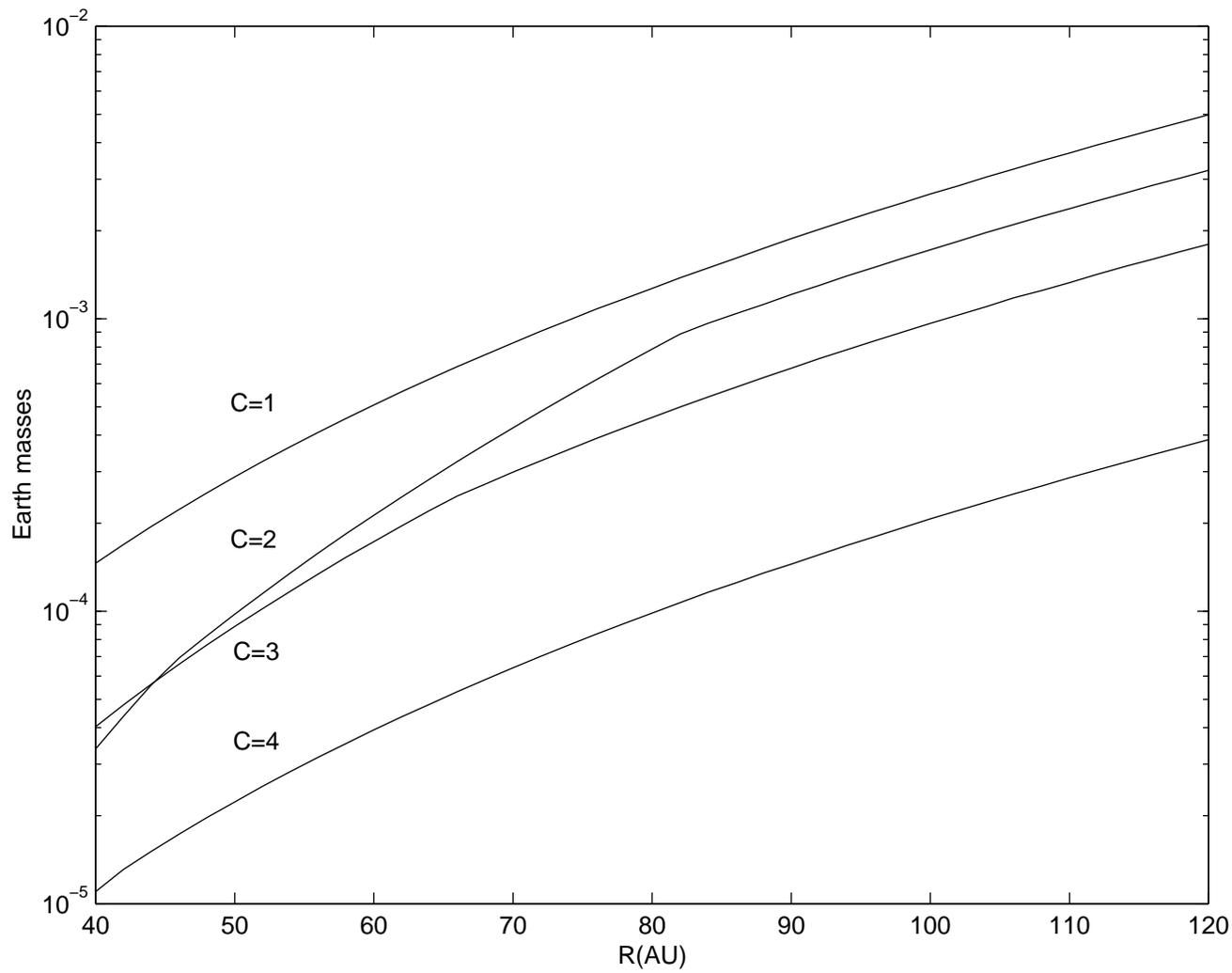
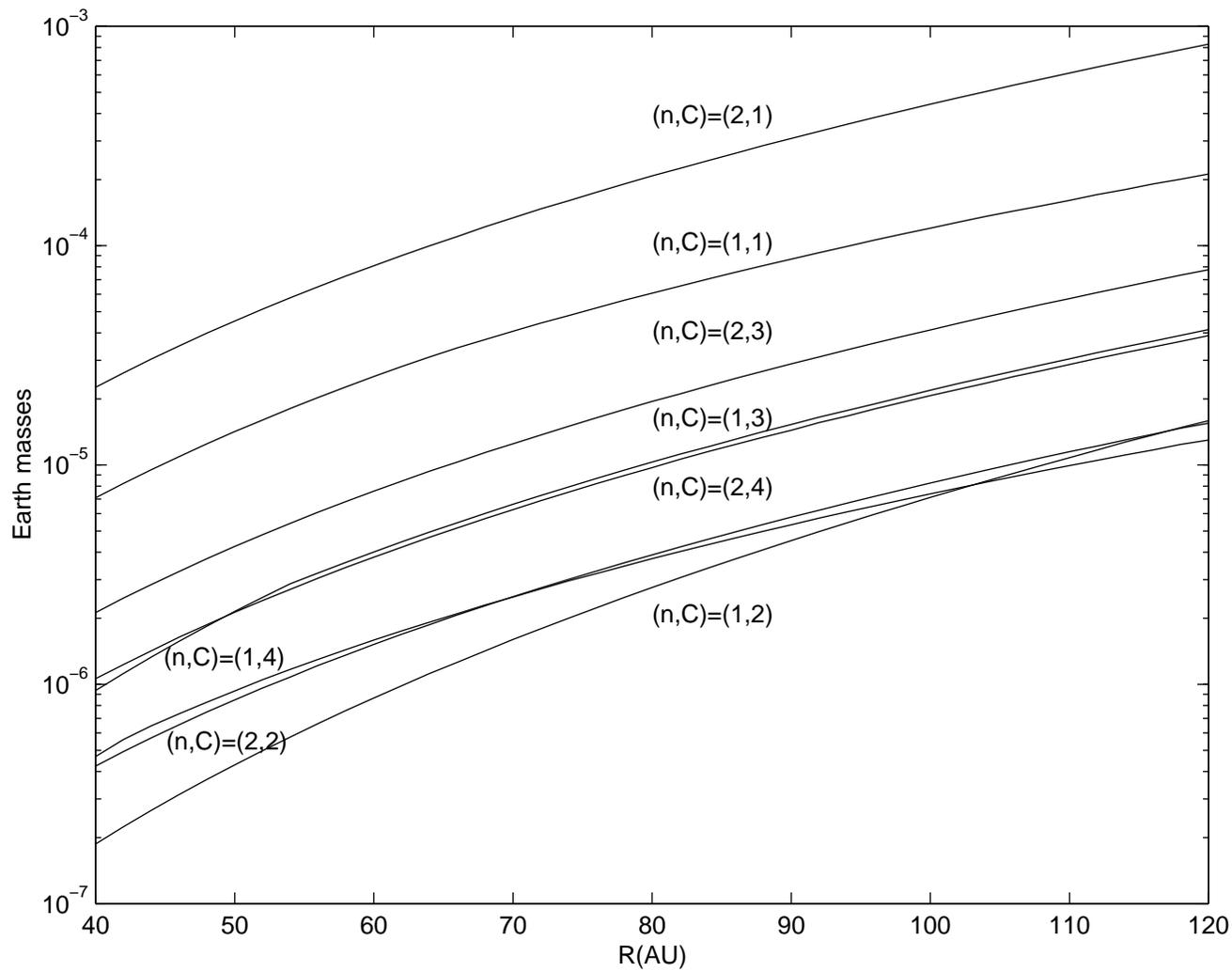
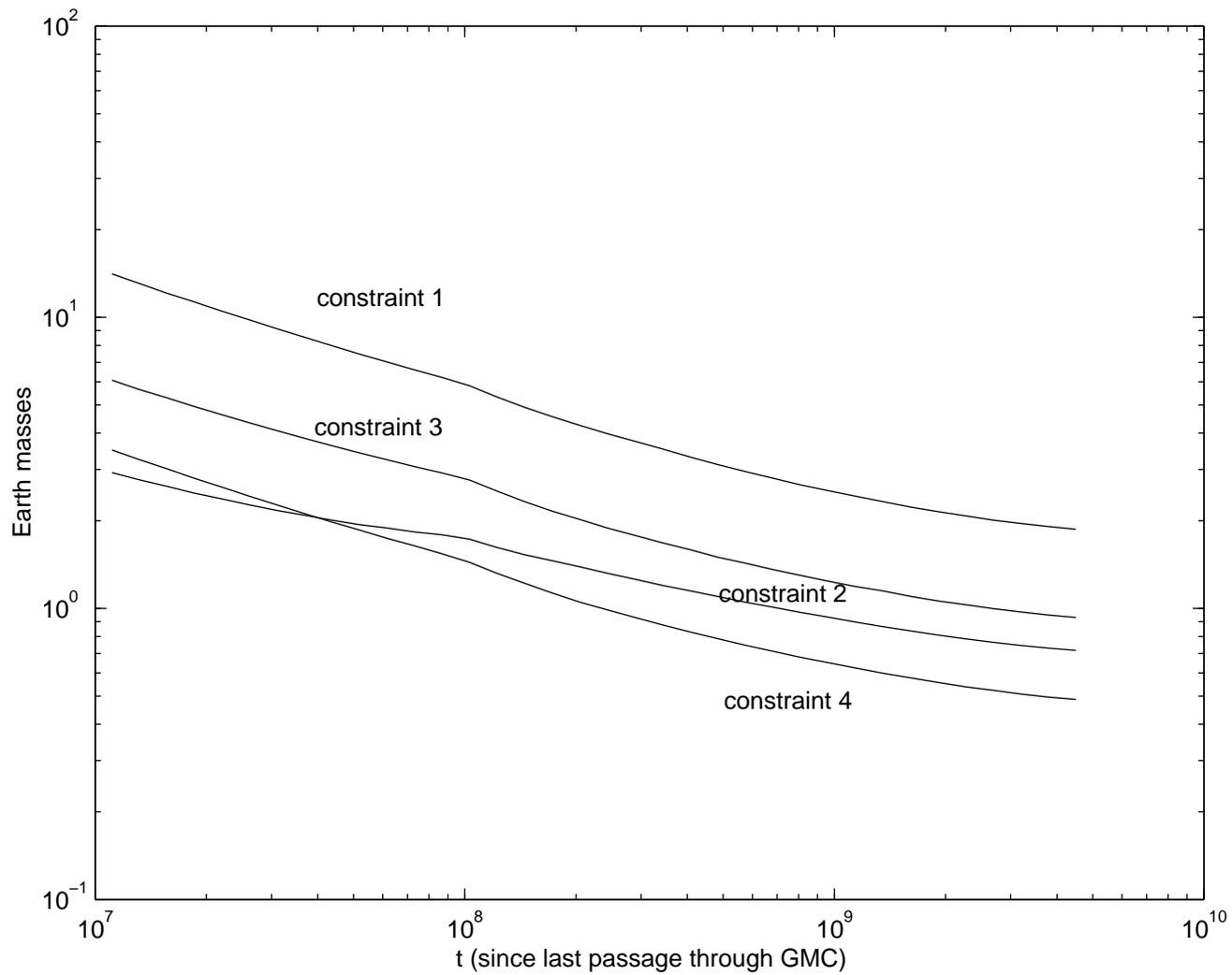
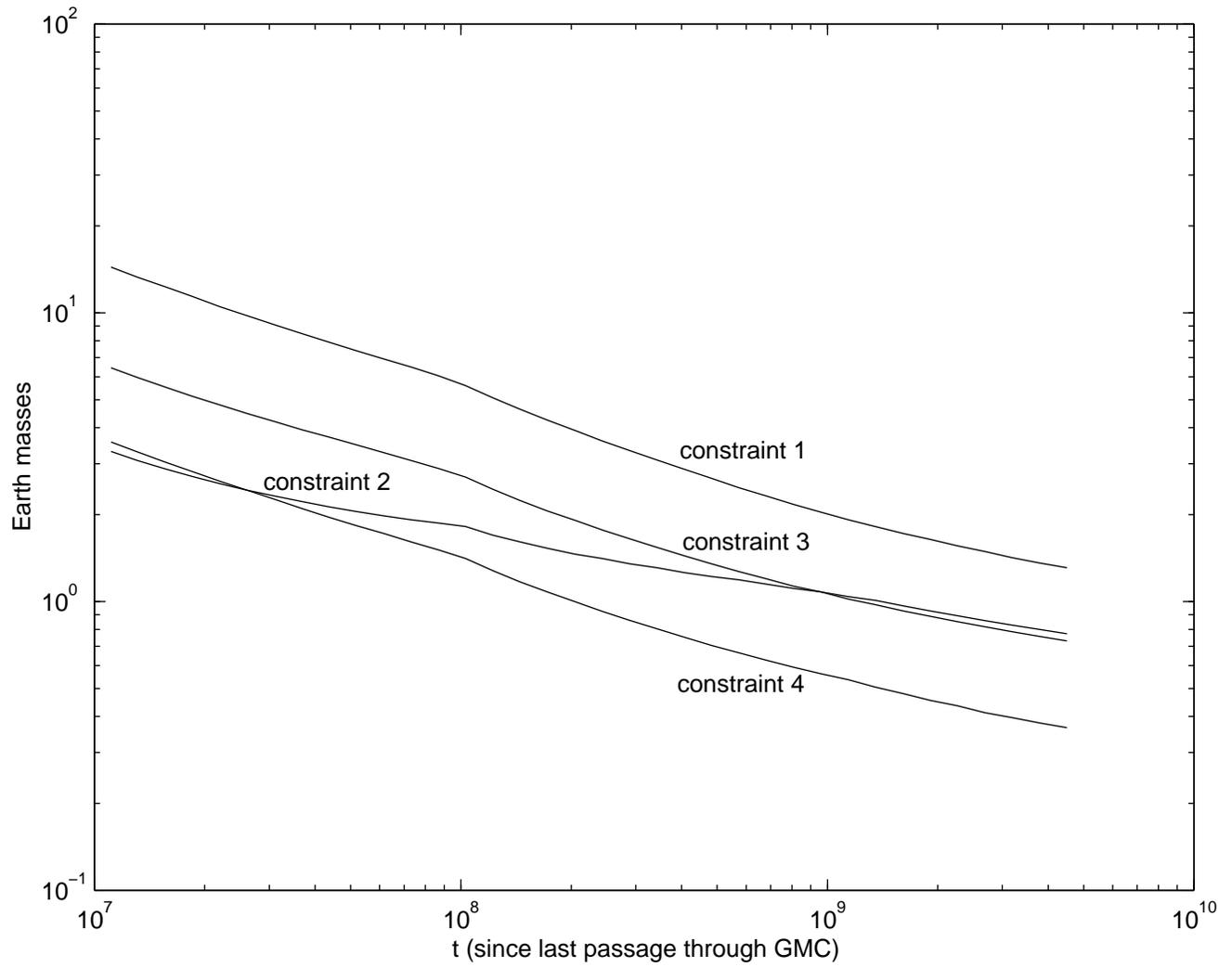
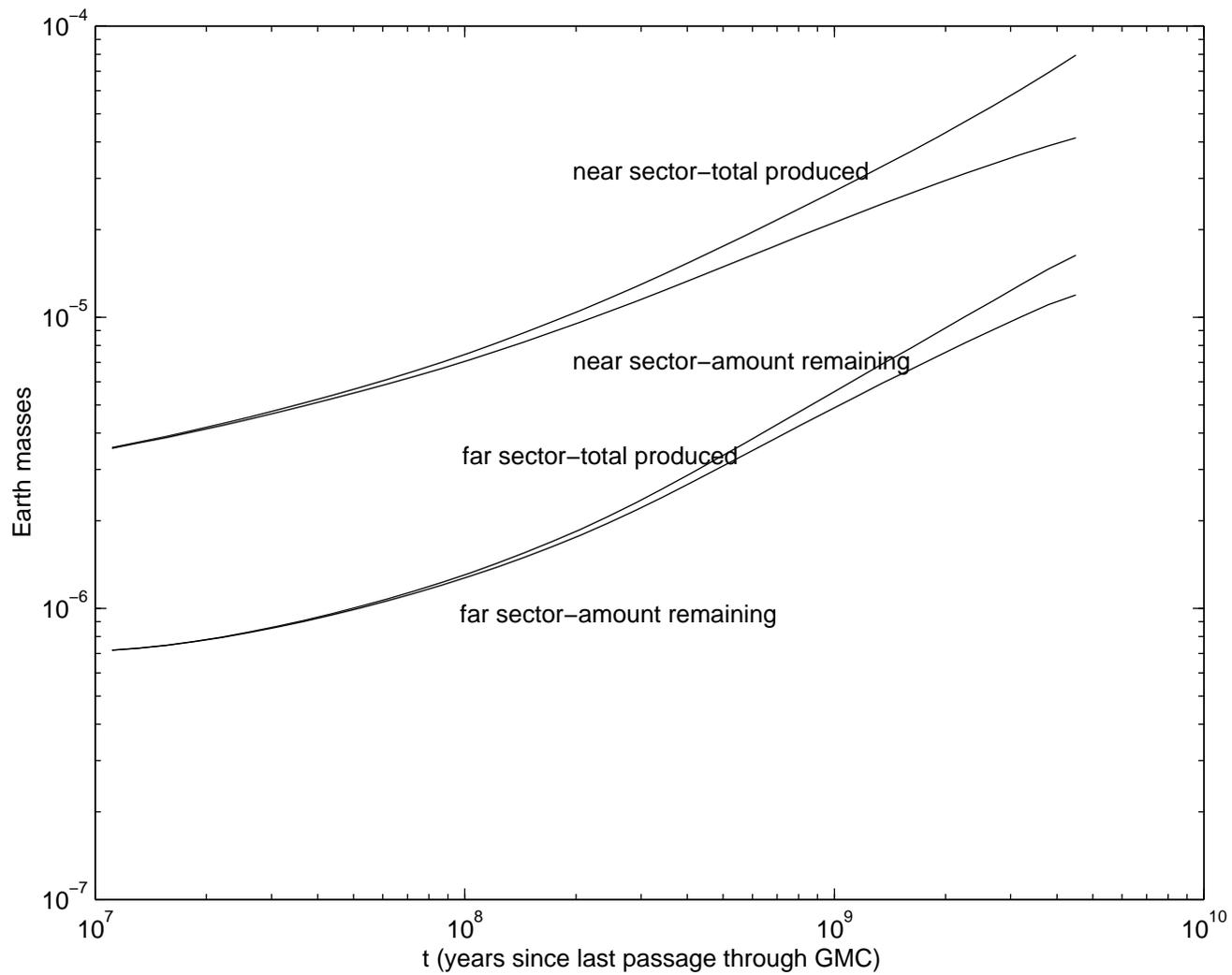
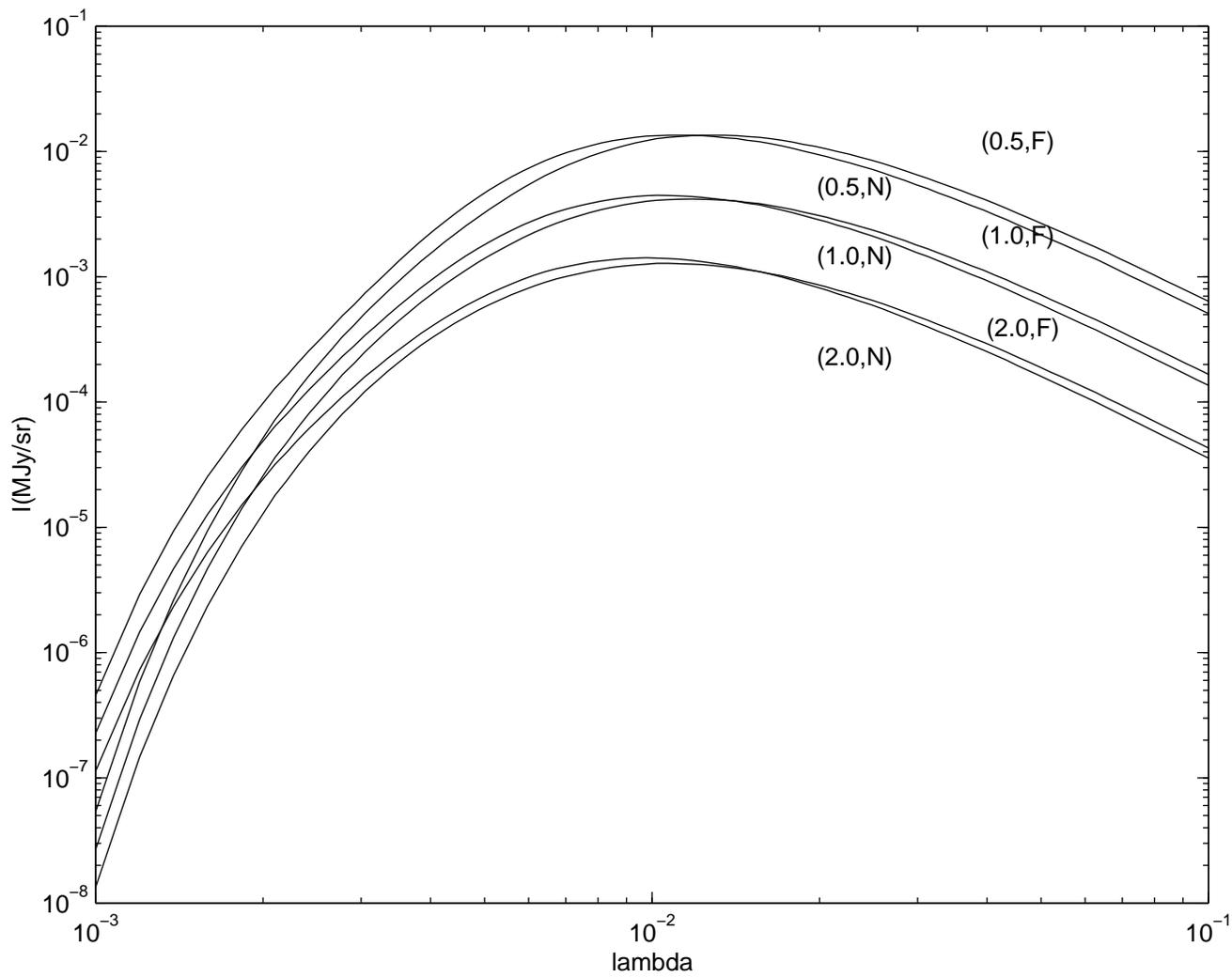
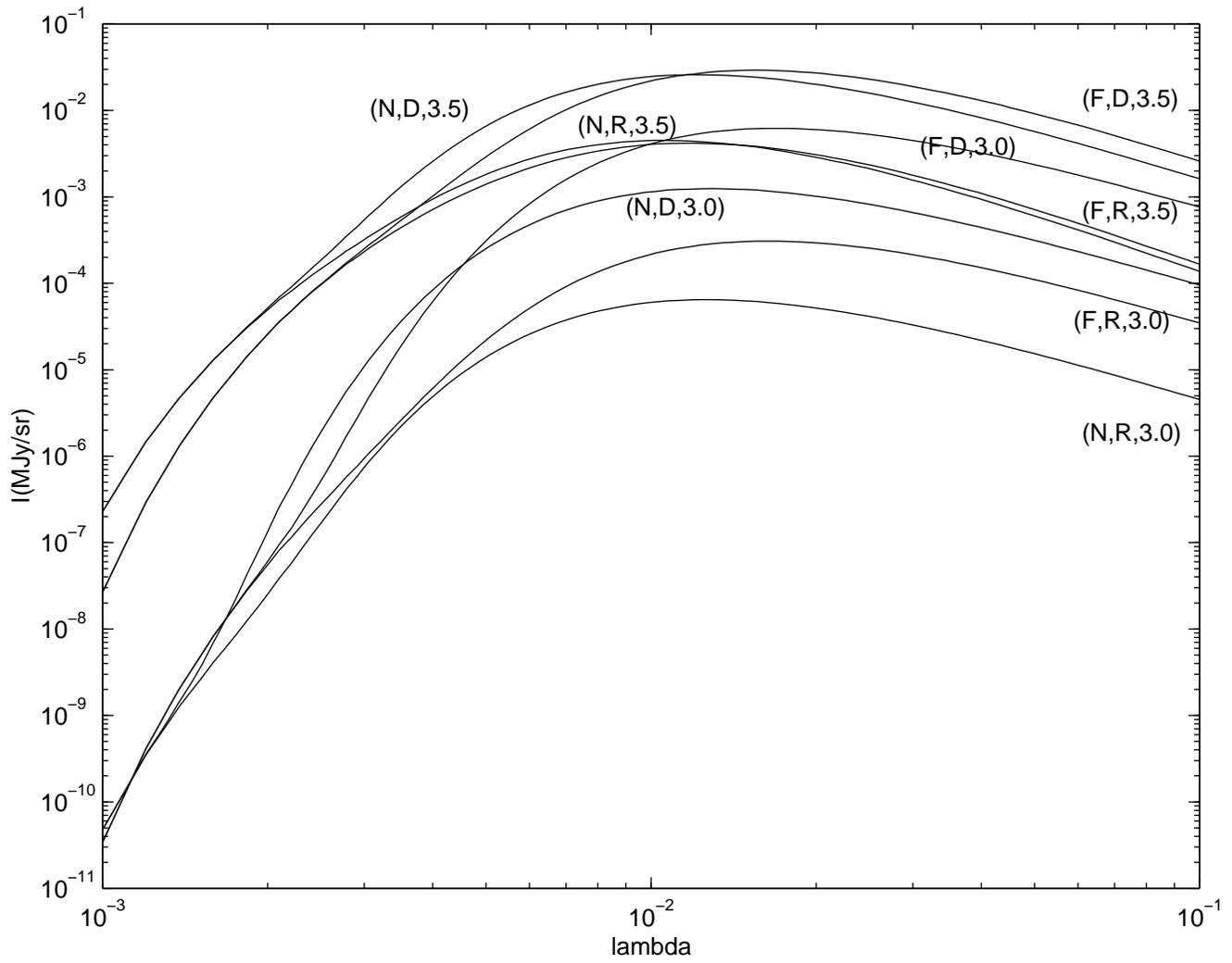
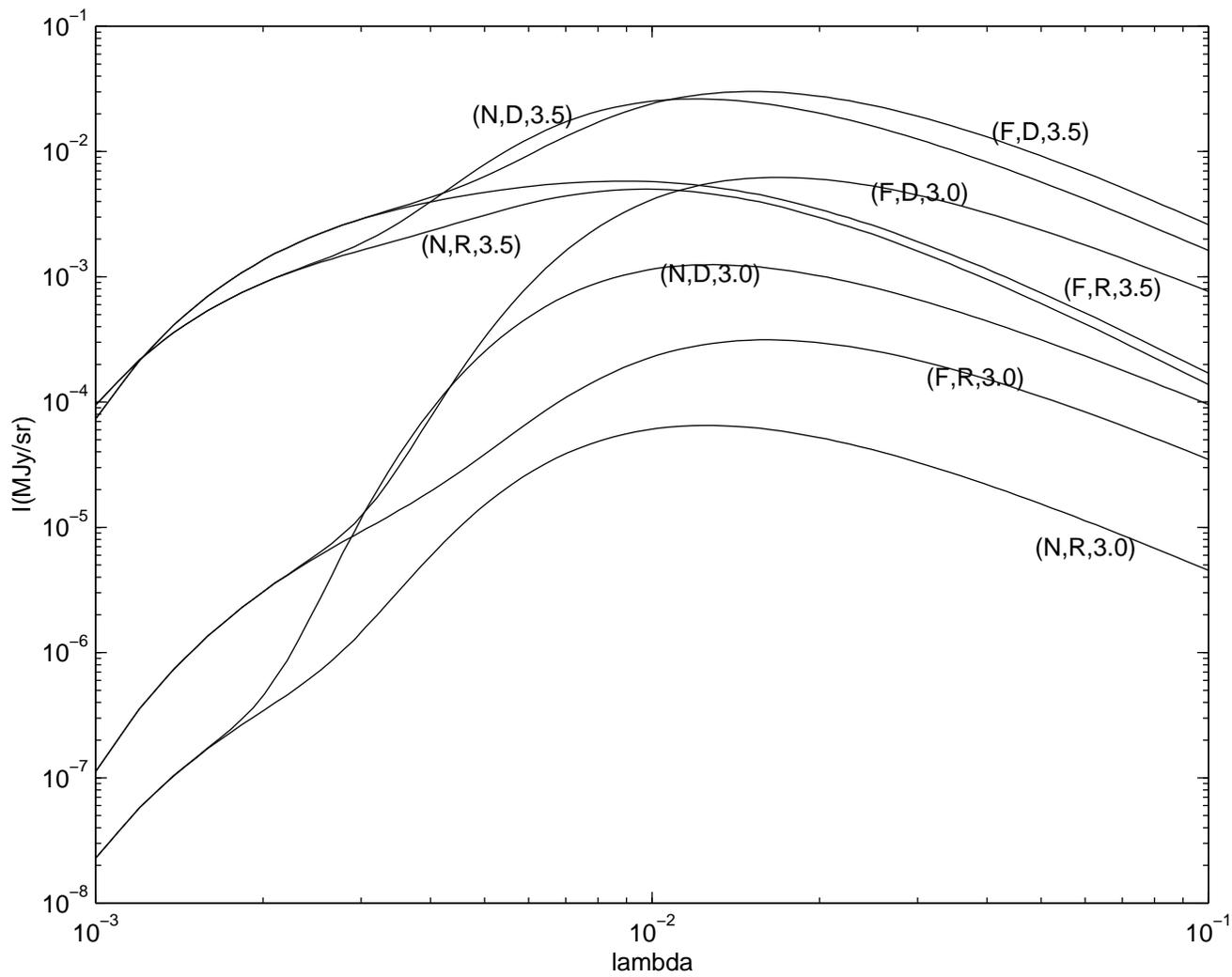
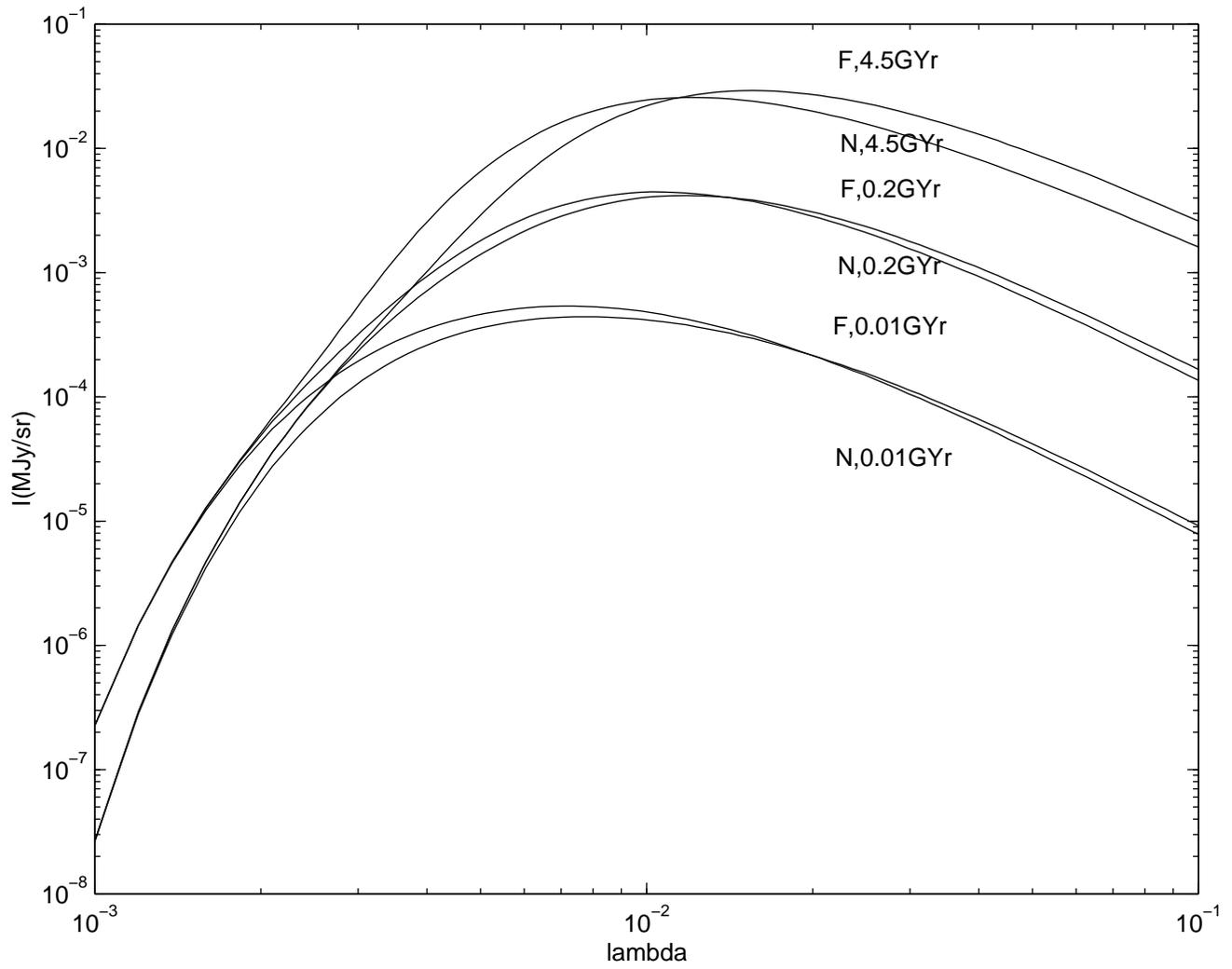
\end{document}